\documentclass[reprint,twocolumn,showpacs,preprintnumbers,amsmath, aps,pre,amssymb]{revtex4-1}
\usepackage{amsmath}
\usepackage{dsfont}
\usepackage{algorithm}
\usepackage{algorithmic}
\usepackage{array}
\usepackage{graphicx}
\usepackage{color}

\newcommand{\vol}{\nu}%\mathbb{D}}
\newcommand{\cut}{E}%{\mathbb{C}}

\newcommand{\NCut}{\mathcal{N}(S)}
\newcommand{\NCutreweighted}{\widetilde{\mathcal{N}}(S)}
\newcommand{\RCut}{\mathcal{R}(S)}
\newcommand{\RCutreweighted}{\widetilde{\mathcal{R}}(S)}
\newcommand{\lambdamax}{\lambda_{max}}

\newcommand{\remove}[1]{}

\begin{document}

\preprint{APS/123-QED}

\title{Spectral Clustering with Epidemic Diffusion}

\author{Laura M. Smith$^1$}

\author{Kristina Lerman$^2$}

\author{Cristina Garcia-Cardona$^3$}

\author{Allon G.\ Percus$^3$}

\author{Rumi Ghosh$^4$}
\affiliation{%
1. California State University, Fullerton, California\\
2. USC Information Sciences Institute, Marina del Rey, California 90292 \\
3. Claremont Graduate University \\
4. HP Labs, Page Mill Road, Palo Alto, California
}%

\date{\today}% It is always \today, today,
             %  but any date may be explicitly specified

\begin{abstract}
Spectral clustering is widely used to partition graphs into distinct modules or communities. Existing methods for spectral clustering use the eigenvalues and eigenvectors of the graph Laplacian, an operator that is closely associated with random walks on graphs. We propose a new spectral partitioning method that exploits the properties of epidemic diffusion. An epidemic is a dynamic process that, unlike the random walk, simultaneously transitions to all the neighbors of a given node.  We show that the replicator, an operator describing epidemic diffusion, is equivalent to the symmetric normalized Laplacian of a reweighted graph with edges  reweighted by the eigenvector centralities of their incident nodes. Thus, more weight is given to edges connecting more central nodes. We describe a method that partitions the nodes based on the componentwise ratio of the replicator's second eigenvector to the first, and compare its performance to traditional spectral clustering techniques on synthetic graphs with known community structure. We demonstrate that the replicator gives preference to dense, clique-like structures, enabling it to more effectively discover communities that may be obscured by dense intercommunity linking.

\end{abstract}

% Check
\pacs{05.45.Xt, 89.75.Hc, 89.75.-k, 89.65.Ef, 89.75.Fb, 02.10.Ud}% PACS, the Physics and Astronomy
                             % Classification Scheme.

\maketitle

\section{Introduction}

Graph partitioning is used in many applications, including community detection~\cite{Leskovec08www}, image segmentation~\cite{ShiMalik00}, and data mining~\cite{Bertozzi2012}, where it is necessary to partition a graph into modules or clusters of similar, or similarly behaving, nodes.
Spectral partitioning uses the eigenvectors associated with the $k$ smallest eigenvalues of the graph Laplacian matrix (or its normalized version)  to partition the graph into $k$ clusters~\cite{Chung:Spectral:97,Ng:2001:NIPS,Spielman07,spectral-tutorial}.
%Spectral partitioning performs well in many settings by solving a linear problem that is easy to implement and can efficiently segment very large graphs.

Existing methods for spectral partitioning are closely associated with random walks on graphs. A random walk is a stochastic dynamic process where transitions take place from a node to a random neighbor of that node, and it is described by the (normalized) graph Laplacian.
The existence of a good partition implies that random walks take a long time to reach a stationary distribution on the graph~\cite{Jerrum88,ShiMalik00}, because they spend a long time within a module and seldom pass between modules~\cite{Rosvall08}. This forms a basis for objective functions used to select which edges to cut so as to partition the graph, such as normalized cut and conductance, though these functions have trouble partitioning real-world graphs where many inter-module edges obscure the underlying structure~\cite{Leskovec08www}.
% modularity maximization
%Physicists more familiar with an alternative objective function - modularity - used to partition the graph into communities. Rather than cut edges, modularity groups nodes so as to maximize ...

Epidemic diffusion is another type of dynamic process on a graph. An epidemic undergoes transitions simultaneously to all the neighbors of a given node, rather than a single neighbor, and is often used to model the spread of a virus or an innovation through a social network~\cite{Anderson91,Rogers03}. Recently, Lerman and Ghosh introduced the replicator matrix~\cite{Lerman12pre}, an analog of the graph Laplacian, to describe epidemic diffusion on graphs. They used the replicator to simulate dynamics of synchronization in a network of oscillators, showing that oscillators coupled via epidemic diffusion synchronize into different structures than oscillators coupled via random walk-like diffusion.

We propose a method for spectral graph partitioning based on epidemic diffusion. First, we show that the replicator is equivalent to the symmetric normalized Laplacian of a reweighted graph, where new edge weights are the product of old edge weights and the eigenvector centralities of the two end points. The eigenvector centrality~\cite{Bonacich01} of a graph is given by the eigenvector corresponding to the largest eigenvalue of the adjacency matrix. Therefore, edges linking central nodes are given a higher weight by the reweighting scheme.

The equivalence between the replicator and symmetric normalized Laplacian of a reweighted graph allows us to exploit well-known relationships   between spectral clustering and graph partitioning.
To use the replicator for spectral partitioning, we give a computationally efficient procedure that orders nodes based on the componentwise ratio  of the second to first eigenvectors and selects a partition that minimizes a quality function computed on the reweighted graph. This tends to preserve dense structures, since edges linking more central nodes in such dense clusters are less likely to be cut.

We study the performance of the proposed spectral partitioning method using synthetic graphs with known community structure. We demonstrate that spectral clustering based on epidemics leads to a better recovery of ground truth communities than traditional methods based on the graph Laplacian, especially in graphs that are more challenging because of the presence of many edges between clusters.
Our work suggests that epidemic diffusion can be a useful probe of graph structure, as it can illuminate properties of graphs that are distinct from those found by methods based on the random walk.

\section{Spectral Clustering}
\label{sec:background}
% Clustering
% Community detection

An unweighted graph $G = (V,E)$, with  vertices (or nodes) $V$ and edges (or links) $E$,  can be represented by a $|V| \times |V|$ adjacency matrix $\boldsymbol{A}$, with $A_{ij}= 1$ if the edge $(i,j) \in E$, and $A_{ij}= 0$ otherwise.  By convention $A_{ii}=0$. We consider undirected graphs, where $A_{ij}=A_{ji}$. The degree of node $i$ is defined as the number of edges incident on it, $d_i= \sum_j {A_{ij}}$. Other useful constructs are $\boldsymbol{D}$, a diagonal degree matrix where $D_{ii}=d_i$, and the identity matrix $\boldsymbol{I}$.

\subsection{Graph Laplacian and Spectral Clustering}
The graph Laplacian matrix is defined as $\boldsymbol{L}=\boldsymbol{D}-\boldsymbol{A}$. The eigenvalues and eigenvectors of $\boldsymbol{L}$ capture many properties of the graph. In the simplest case, if the graph has $k$ disjoint components, the $k$ smallest eigenvalues of $\boldsymbol{L}$ are zero, and the associated eigenvectors are indicator functions assigning nodes to their respective cluster or community~\cite{spectral-tutorial}. Even if the $k$ smallest eigenvalues are not all zero, their corresponding eigenvectors can be used to partition nodes into $k$ clusters by projecting these nodes onto a subspace  of the first $k$ eigenvectors and using standard clustering techniques such as $k$-means~\cite{Ng:2001:NIPS,ShiMalik00}.
% to partition nodes into $k$ clusters
The simplest spectral clustering method, spectral bisection, partitions nodes based on the values of the second eigenvector $\boldsymbol{v}$ of the adjacency matrix or the graph Laplacian. A splitting value $c$ is used to divide the nodes into different clusters based on whether  $\textbf{v}_i < c$ or $\textbf{v}_i \geq c$~\cite{Spielman07}.  A range of splitting values have been used, including zero, the median value within the vector, the largest gap, and the value producing the best ratio cut, best conductance~\cite{Vempala2009}, or another measure.

In practice, normalized versions of the graph Laplacian produce better results in spectral clustering applications~\cite{Ng:2001:NIPS,Bertozzi2012}. Two examples are the symmetric normalized Laplacian $\boldsymbol{L}_s=\boldsymbol{I}-\boldsymbol{D}^{-1/2} \boldsymbol{A} \boldsymbol{D}^{-1/2}$ and the random walk Laplacian $\boldsymbol{L}_{rw}=\boldsymbol{I}-\boldsymbol{D}^{-1}\boldsymbol{A}$, so named because the matrix of transition probabilities for a random walk on a graph is given by $\boldsymbol{D}^{-1}\boldsymbol{A}$.

\subsection{Graph Cuts and Their Quality Measures}
Intuitively, a cluster is a set of nodes $S \subset V$ that are more tightly connected to each other than to nodes outside of the cluster. We  use $\bar{S}=V\setminus S$ to denote the complement of $S$, which consists of nodes that are not in $S$. In order to bisect the graph into two disjoint clusters, one typically wants to minimize the number of cut edges between clusters,
$$ \cut(S,\bar{S})=\sum_{i \in S, j \in \bar{S}}A_{ij},$$
while maximizing cluster size, which may be measured by the number of nodes it contains, $|S|$, or the  sum of the degrees of the nodes in the set, $\vol(S)=\sum_{i \in S} {d_i}$, also called volume of the set.

Several functions have been proposed for measuring the quality of a graph cut. The best known of these are ratio cut $\RCut$ and normalized cut $\NCut$:
\begin{eqnarray}
%\label{eq:conductance}
%    \phi(S) & = & \frac{cut(S,\bar{S})}{{\min(vol(S),vol(\bar{S}))}} \\
    \label{eq:ratiocut}
    \RCut & = & \left(\frac{1}{|S|}+\frac{1}{|\bar{S}|}\right) \cut(S,\bar{S})\\
    \label{eq:ncut}
    \NCut & = & \left(\frac{1}{\vol(S)}+\frac{1}{\vol(\bar{S})}\right)\cut(S,\bar{S}).
\end{eqnarray}

There is a relationship between graph cuts and spectral clustering. Deciding which edges to cut to optimize any of these quality functions is an NP-complete problem. Spectral clustering solves a relaxation of the problem, where the discrete indicator variables that assign nodes to clusters become continuous.  Although in general there are no useful bounds for the approximation produced by this relaxation~\cite{spectral-tutorial}, in practice it often provides a simple and effective clustering method.  Solutions to the relaxed optimization problem are given by the second eigenvector of the graph Laplacian $\boldsymbol{L}$ or the normalized graph Laplacian $\boldsymbol{L}_s$~\cite{Spielman07}.  Relaxing ratio cut leads to spectral clustering using $\boldsymbol{L}$, while relaxing normalized cut leads to spectral clustering using $\boldsymbol{L}_s$~\cite{ShiMalik00,spectral-tutorial}. Such relaxation methods have also been applied productively to the popular modularity maximization method for community detection~\cite{Newman06pnas,Fortunato10community}. By analogy with spectral bisection~\cite{Spielman07}, the leading eigenvector approach assigns nodes to clusters based on the sign of the components of the leading eigenvector of the modularity matrix.

\subsection{Spectral Clustering and Random Walks}
There exists a further relationship between spectral clustering, the partition quality function, and properties of random walks. A random walk on a graph is a stochastic process where transitions take place to a randomly chosen neighbor of a given node. Cluster properties of the graph can be expressed in terms of the transition matrix $\boldsymbol{D}^{-1}\boldsymbol{A}$~\cite{Lovasz93} of a random walk. Spectral clustering finds a partition such that a random walk stays within the same cluster for a long time and seldom jumps between clusters~\cite{ShiMalik00,Rosvall08}. Therefore, the presence of a good partition (low normalized cut value) implies  that it will take a random walk a long time to reach its equilibrium distribution.

\section{Epidemic Diffusion on Graphs}
\label{sec:replicator}
An epidemic is a dynamic process that simultaneously undergoes transitions to every neighbor of the current node. Epidemics are used to model the spread of disease~\cite{Hethcote00} and innovation~\cite{Rogers03} in social networks. Epidemics differ from random walks in important ways.  First, rather than choosing a single neighbor to transition to or ``infect'' as the random walk does, an epidemic will attempt to ``infect'' every neighbor of a node.  In a random walk, the probability of finding the walker in a given location is a conserved quantity that diffuses through the graph, and the random walk transition matrix is a stochastic matric.  Epidemics, on the other hand, replicate themselves with each successful transmission, without following a conservation law~\cite{Lerman12pre}.

Lerman and Ghosh~\cite{Lerman12pre} introduced the replicator operator $\boldsymbol{R}=\lambdamax\boldsymbol{I}- \boldsymbol{A}$ to describe dynamics of synchronization in a network of nodes coupled via epidemic diffusion.  Here $\lambdamax$ is the largest eigenvalue of $\boldsymbol{A}$, also known as the epidemic threshold~\cite{Wang03}.  In this system, a dynamic variable $u_i$ associated with node $i$ can change its value based on the values of its neighbors according to:
\begin{equation}
\label{eq:evolution}
 \frac{d\boldsymbol{u}}{dt}=-\boldsymbol{R}\boldsymbol{u},
\end{equation}
\noindent where $\boldsymbol{R}$ replaces the Laplacian used in the
analogous heat equation that gives the (diffusive) evolution of a random
walk on a graph~\cite{Chung07pnas}.
By construction, the replicator has a steady state
given by $\boldsymbol{\theta}$, the eigenvector of $\boldsymbol{A}$ associated with $\lambdamax$: $\boldsymbol{A} \boldsymbol{\theta}=\lambdamax \boldsymbol{\theta}$.    $\boldsymbol{\theta}$ is also known as the \emph{eigenvector centrality}~\cite{Bonacich01}, and was introduced by Bonacich to explain the  importance of actors in a social network based on the importance of the actors to which they were connected.

Clusters of nodes with similar values of the dynamic variable $\boldsymbol{u}$ emerge as the system of coupled nodes evolves towards the steady state~\cite{Lerman12pre}.
This motivates a community detection method with nodes classified
according to the rate of convergence to their steady-state values.  For
large time $t$, we approximate the solution to Eq.~\ref{eq:evolution}
using the two leading eigenvectors $\boldsymbol{\theta}$ and
$\boldsymbol{\psi}$ of $\boldsymbol{R}$,
\begin{eqnarray*}
u_i(t) &\approx& c_1 \theta_i + c_2 e^{-\lambda_2 t} \psi_i \\
&=& c_1 \theta_i \left[ 1+ \frac{c_2}{c_1} e^{-\lambda_2 t}
\frac{\psi_i}{\theta_i}\right],
\end{eqnarray*}
where $c_1$ and $c_2$ are constants, and $\lambda_2$ is the second smallest
eigenvalue of $\boldsymbol{R}$ associated with eigenvector
$\boldsymbol{\psi}$, guaranteed to be nonzero if the graph is connected.
Therefore, convergence depends on
$\psi_i/\theta_i$, the componentwise ratio of the second to first
eigenvectors.   Note that  eigenvectors of $\boldsymbol{R}$ corresponding to $\boldsymbol{R}$'s two smallest eigenvalues are the same as the eigenvectors of $\boldsymbol{A}$  corresponding to $\boldsymbol{A}$'s two largest eigenvalues.

\subsection{Replicator as the Symmetric Normalized Laplacian of a Reweighted Graph}
In a social network, one might expect nodes of high ``importance''
to attract other nodes, resulting in communities forming
around nodes with large eigenvector centrality values $\theta_i$.
In this section we propose a modification of our graph, converting the
unweighted network into a weighted one where weights are given by the
product of the eigenvector centralities of an edge's end points:
$\tilde{A}_{ij}=A_{ij}\theta_i \theta_j$.  Moreover, we show that the
replicator on the unweighted graph given by $\boldsymbol{A}$ is in fact exactly equivalent to the
symmetric normalized Laplacian of the reweighted graph given by $\widetilde{\boldsymbol{A}}$.

In the reweighted graph, the degree of node $i$ is given by
\begin{equation*}
\label{eq:degree}
\tilde{d}_i=\sum_j{A_{ij}\theta_i
\theta_j}=\theta_i\sum_j{A_{ij}\theta_j}=\lambdamax\theta_i^2.
%=\lambdamax \: \theta_i \: \theta_j \: \delta_{ij}.
\end{equation*}
For convenience, define
$\boldsymbol{\Theta}$ as the diagonal matrix whose elements are the
components of eigenvector $\boldsymbol{\theta}$, i.e.,\
$\boldsymbol{\Theta}_{ii}$.  Then, from $\tilde{A}_{ij}$ and
$\tilde{d}_i$ above,
% \begin{equation*}
% 	\lambdamax \: \theta_i \: \theta_j \: \delta_{ij} = \lambdamax \: \left ( \boldsymbol{\Theta} \; \boldsymbol{\Theta} \right )_{ii}
% \end{equation*}
%
%
% Thus, the degree matrix of the reweighted graph is:
\begin{equation}
\label{eq:reweight}
	\widetilde{\boldsymbol{A}} = \boldsymbol{\Theta}\boldsymbol{A}\boldsymbol{\Theta}
        \quad\mathrm{and}\quad
	\widetilde{\boldsymbol{D}} = \lambdamax \: \boldsymbol{\Theta}^2.
\end{equation}
We can now write the symmetric normalized Laplacian of the reweighted
graph:
\begin{eqnarray*}
\widetilde{\boldsymbol{L}}_{\boldsymbol{s}} &=& \boldsymbol{I} -
\widetilde{\boldsymbol{D}}^{-1/2}
\widetilde{\boldsymbol{A}}\widetilde{\boldsymbol{D}}^{-1/2} \\
		& = & \boldsymbol{I} - \left (
\frac{1}{\sqrt{\lambdamax}} \; \boldsymbol{\Theta}^{-1} \right ) \;
\boldsymbol{\Theta}\boldsymbol{A}\boldsymbol{\Theta}\; \left ( \frac{1}{\sqrt{\lambdamax}} \; \boldsymbol{\Theta}^{-1} \right ) \\
%		& = & \boldsymbol{I} - \frac{1}{\lambdamax} \; \boldsymbol{\Theta}^{-1} \; \widetilde{\boldsymbol{A}} \; \boldsymbol{\Theta}^{-1} \\
%		& = & \boldsymbol{I} - \frac{1}{\lambdamax} \; \boldsymbol{\Theta}^{-1} \; \boldsymbol{\Theta} \; \boldsymbol{A} \; \boldsymbol{\Theta} \; \boldsymbol{\Theta}^{-1} \\
		& = & \boldsymbol{I} - \frac{1}{\lambdamax} \; \boldsymbol{A} \\
		& = & \frac{1}{\lambdamax} \; \boldsymbol{R}.
\end{eqnarray*}
\noindent Hence,
$\boldsymbol{R}=\lambdamax\widetilde{\boldsymbol{L}}_{\boldsymbol{s}}$.

The equivalence between epidemics and the diffusive process of random walks is at first surprising.  Diffusive processes conserve the total amount of the substance diffusing, whereas no such conservation law holds for epidemics~\cite{Lerman12pre}.
% The equivalence between random walks and epidemics is at first surprising. In a random walk, the probability to find the walker on any node of the graph is conserved, while epidemics are non-conservative~\cite{Lerman12pre}.
The intuition for the equivalence of the two processes is the following. A node's eigenvector centrality gives the number of paths connecting it to all other nodes in the graph~\cite{Ghosh11physrev}; hence, the product of eigenvector centralities of a pair of nodes captures how much of the substance is newly created when the epidemic follows the edge linking the pair. By encoding the amount of non-conservation in edge reweighting, this scheme allows the epidemic to be reduced to diffusion.

\subsection{Quality Measure for the Replicator}
The equivalence proved above allows us to exploit the properties of the symmetric normalized Laplacian, along with its relationship to graph partitioning, for epidemic diffusion.
Since the replicator is simply $\boldsymbol{L}_s$  of the reweighted graph $\widetilde{\boldsymbol{A}}$,
spectral clustering using the replicator corresponds to a
relaxation of normalized cut on this reweighted graph.  The appropriate measure for assessing graph cut quality with the replicator is therefore normalized cut on the reweighted graph $\NCutreweighted$.

\subsection{An Illustrative Example}\label{sec:toygraph}
\begin{figure}[!htb]
\centering
\begin{tabular}{cc}
\includegraphics[,width=1.65in]{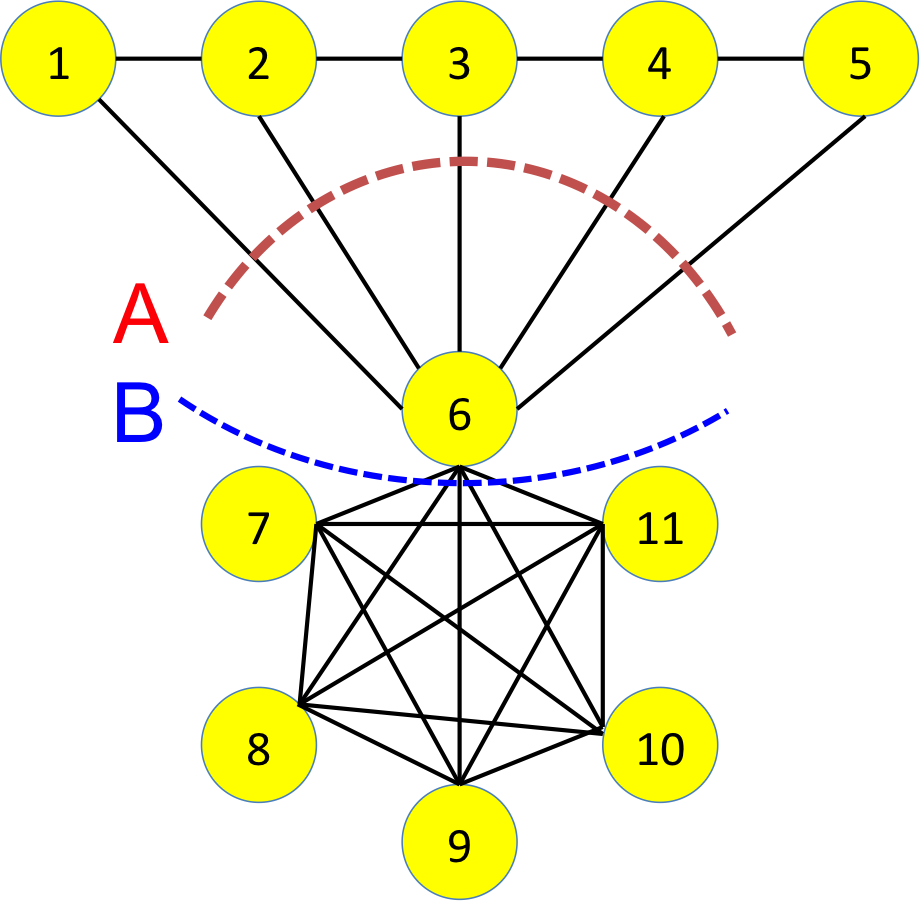}&
\raisebox{2.2cm}
{
\begin{tabular}{ |c|c|c| }
  \hline
Quality& Cut A & Cut B\\
\hline
% Conductance $\phi$  & Original &0.385 & \textbf{0.217}\\
\multicolumn{3}{|c|}{\emph{Original graph}} \\ \hline
$\RCut$  & \textbf{1.83} & \textbf{1.83}\\
$\NCut$ & 0.528& \textbf{0.417}\\
\hline
% Conductance $\phi$  & Reweighted &0.683 & \textbf{0.536}\\
\multicolumn{3}{|c|}{\emph{Reweighted graph}} \\ \hline
$\RCutreweighted$  & \textbf{11.4} & 32.3\\
$\NCutreweighted$ & \textbf{0.747}& 0.778\\
  \hline
\end{tabular}
}
\end{tabular}
\caption{(Color Online)(Left) An example graph.   The possible cuts are shown by the dotted curves A and B. (Right) Quality measures of cuts A and B on the original and reweighted graph.}\label{fig:toy}
\end{figure}

We use a simple example to highlight the differences between traditional graph partitioning and one based on epidemics. Consider the graph in Figure~\ref{fig:toy}, which shows a dense cluster connected through node 6 to a sparsely linked cluster. Such a configuration is common in social networks, where a high-degree hub linking different communities may obscure community boundaries. We expect a good partition  to group node $6$ with other nodes in its clique. However, the cut ($B$) that minimizes normalized cut ($\NCut$) groups node $6$ with nodes $1$--$5$ and assigns nodes $7$--$11$ to the other cluster.  Multiple cuts minimize ratio cut ($\RCut$), including one that groups together nodes $3$--$5$.
\remove{Modularity maximization technique also prefers cut $B$.}

Node $6$ has the highest eigenvector centrality. Furthermore, nodes that belong to the clique have higher centrality values than other nodes.  Consequently, in the {\em reweighted\/} graph, the edges linking node $6$ to the rest of the clique are more ``expensive'' to cut, and nodes $6$--$11$ are grouped together by the preferred cut ($A$) that minimizes both the ratio cut $\RCutreweighted$ and the normalized cut $\NCutreweighted$ on the reweighted graph.  The quality measures of the cuts are shown in the table in Fig.~\ref{fig:toy}.   By giving edges linking central nodes a higher weight, epidemic-based graph partitioning thus preserves dense, clique-like structures. Accordingly, deleting these edges will have the greatest impact on reducing the spread of an epidemic~\cite{Tong:2012}.

\subsection{Efficient Spectral Partitioning}

We now describe an efficient method for spectral clustering using epidemic diffusion based on spectral bisection~\cite{Spielman07}. First, we create
a vector $\boldsymbol{v}$ that is the componentwise
ratio of the second eigenvector $\boldsymbol{\psi}$ to the first
eigenvector $\boldsymbol{\theta}$ of the operator $\boldsymbol{R}$ and
sort its values. Next, we examine all  $N-1$
cuts in this ordering (where $N=|V|$) and pick one corresponding to the partition that minimizes an appropriate quality measure.  The quality measure we use
with $\boldsymbol{R}$ is normalized cut on the reweighted graph ($\NCutreweighted$).
We compare the resulting partition with those
produced by applying an analogous splitting procedure to $\boldsymbol{L}$, with quality
measure $\RCut$, and $\boldsymbol{L}_s$, with quality
measure $\NCut$ (on the original graph).

The proposed optimization procedure is exhaustive, since it tests all $N-1$ possible cuts within the ordering produced by $\boldsymbol{v}$.  It may seem that there would be some loss in accuracy from restricting our search to cuts in a one-dimensional projection, rather than searching over the entire subspace spanned by the first two eigenvectors $\boldsymbol{\theta}$ and $\boldsymbol{\psi}$.  However, it has been observed~\cite{weiss,ShiMalik00} that the componentwise ratio of the second to first eigenvector of $\boldsymbol{L}_s$ is precisely equal to the second eigenvector of the random walk Laplacian $\boldsymbol{L}_{rw}$, whose first eigenvector is a constant vector.  Thus, our algorithm is effective because it is a computationally efficient procedure for finding the best normalized cut in the two-dimensional eigenspace of $\widetilde{\boldsymbol{L}}_{rw}$, i.e., $\boldsymbol{L}_{rw}$ on the reweighted graph.  The advantages of using $\boldsymbol{L}_{rw}$ in spectral clustering are discussed in~\cite{spectral-tutorial}.

\section{Evaluation on Synthetic Graphs}
We use synthetic graphs to gain better insight into the differences between operators $\boldsymbol{L}$, $\boldsymbol{L}_s$, and $\boldsymbol{R}$ and the characteristics of graphs for which different operators find better solutions.
Lancichinetti and Fortunato have proposed an algorithm to generate random graphs with known hierarchical community structure~\cite{Fortunato2009}.
The $N$ nodes are divided into macro communities, which are themselves composed of micro communities, and then edges between nodes are created using mixing parameters $\mu_1$ and $\mu_2$.  The parameter $\mu_1$ designates the fraction of a node's edges that will connect to nodes in a different macro community, and $\mu_2$ gives the fraction of edges that will connect to nodes in a different micro community within the same macro community.  The remaining $\left(1-\mu_1-\mu_2\right)$ fraction of edges link to other nodes within the same micro and macro communities.
These benchmark networks allow us to systematically explore the performance of different spectral clustering approaches.

\begin{figure}[htb]
\centering
\includegraphics[width=1.4in]{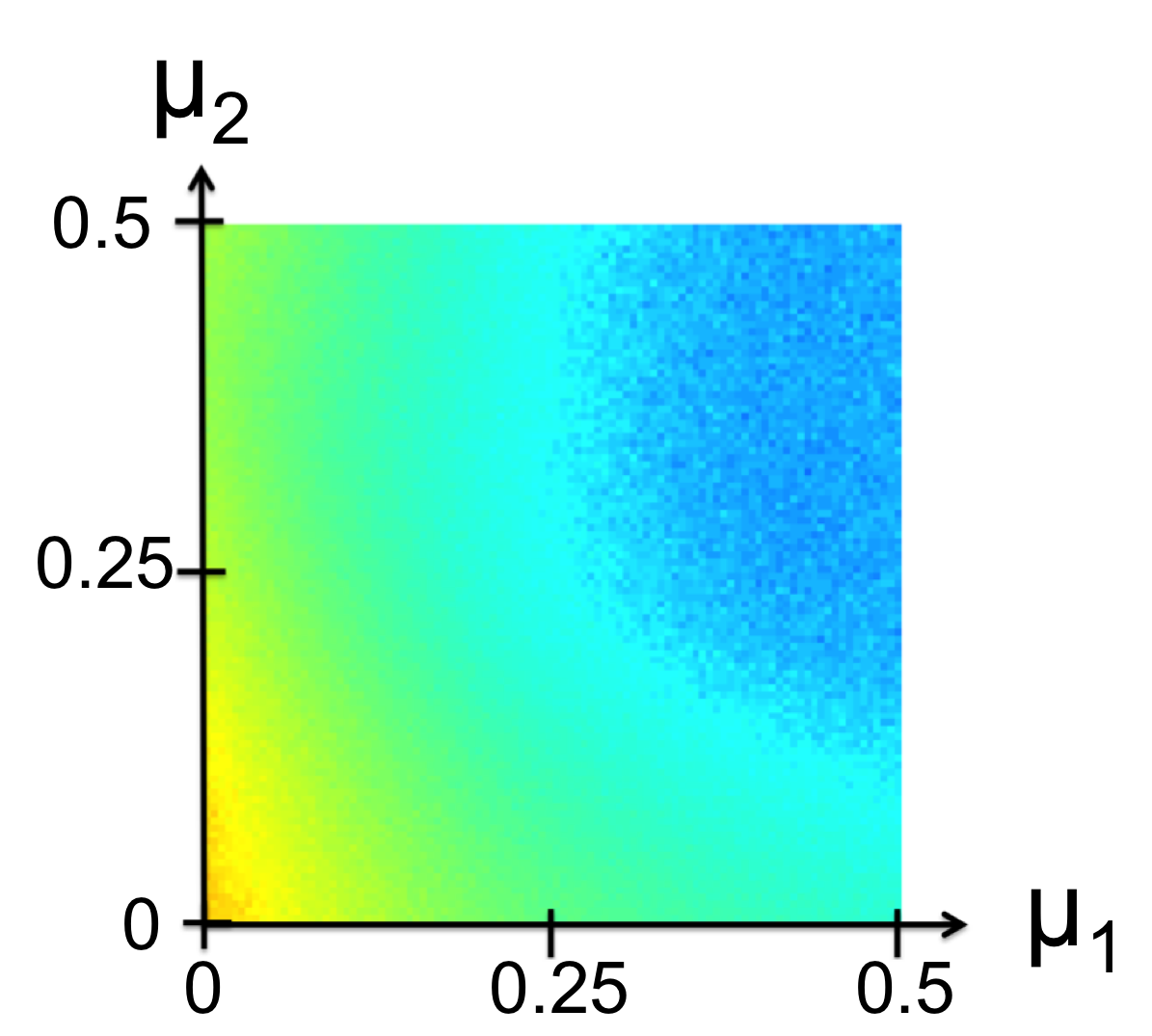}
\includegraphics[width=0.2in, height=1.0in]{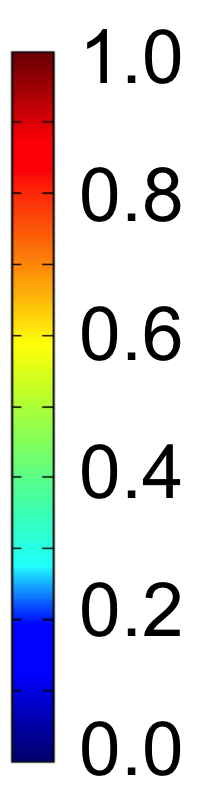}
\includegraphics[width=1.4in]{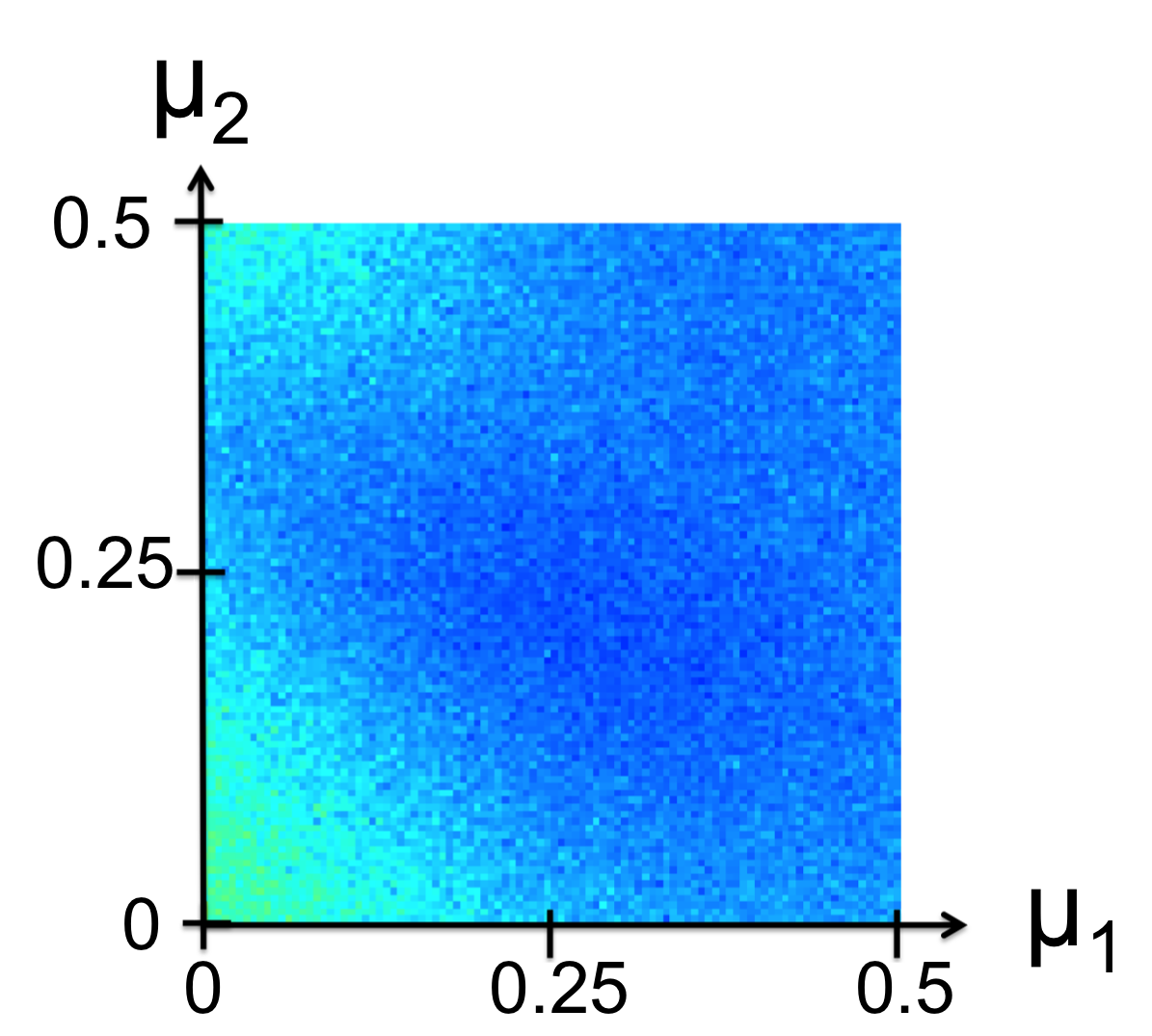}
\includegraphics[width=0.2in, height=1.0in]{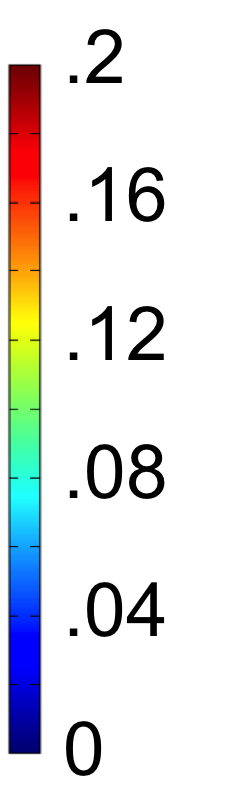}
\caption{Each pixel  represents the mean average clustering coefficient (left) and the standard deviation (right) across 100 runs for fixed  $(\mu_1, \mu_2)$. \label{fig:Aplots}
}
\end{figure}

Using software available on \cite{FortunatoWebsite}, we generated 100 graphs for each set of parameter values.  We took $N=100$ with two macro communities.  We varied $\mu_1$ and $\mu_2$ between $0$ and $0.5$. The average clustering coefficient ranged between  0.23 and  0.6421, suggesting that the synthetic graphs have properties similar to those often found in real world  networks \cite{Watts1999}.

\begin{figure}[!htb]
\centering
\begin{tabular}{@{}c@{}c@{}l@{}}\\
\\
\multicolumn{3}{c}{Laplacian $\boldsymbol{L}$: Ratio Cut} \\
\includegraphics[width=1.6in]{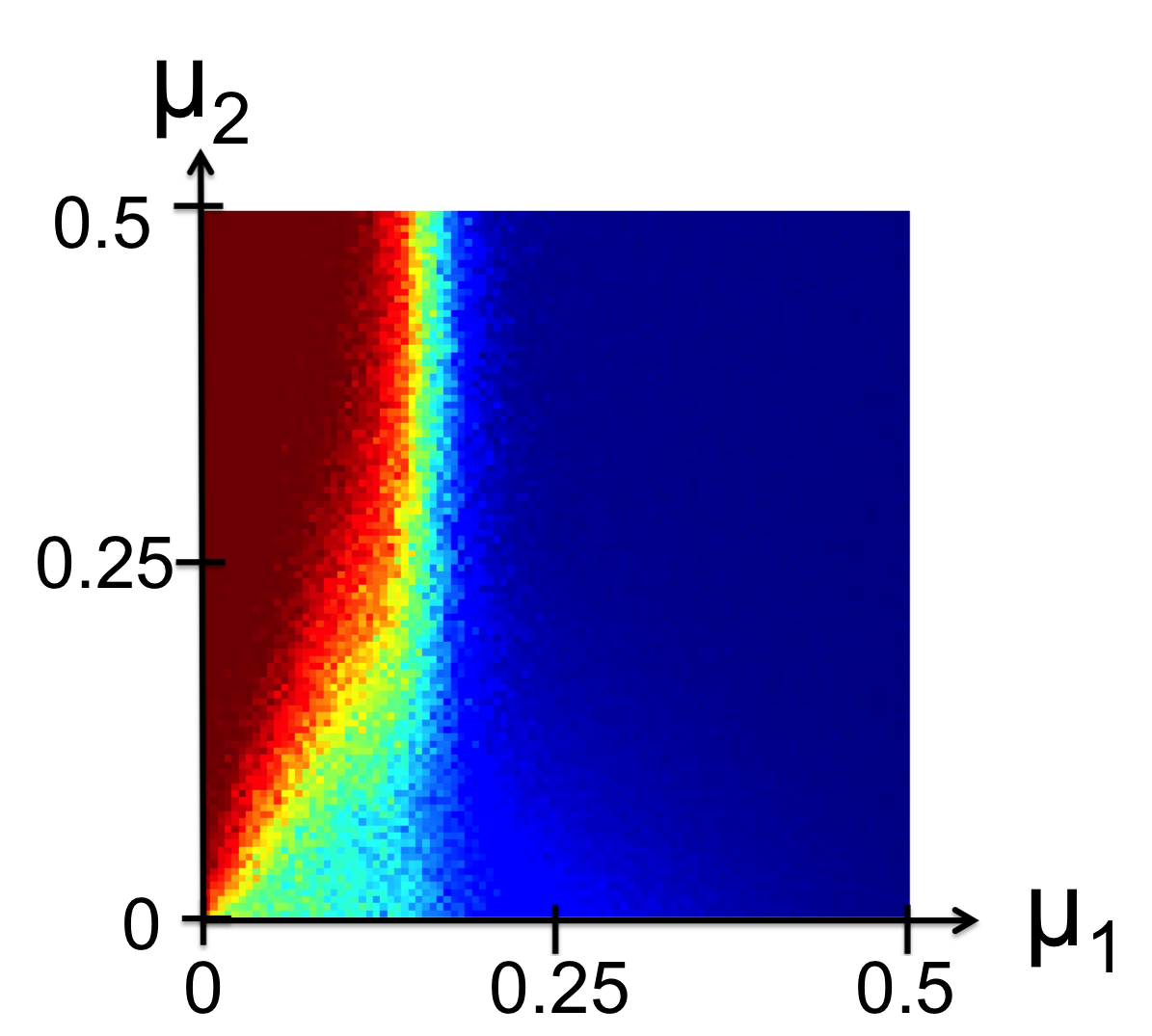} &
\includegraphics[width=1.6in]{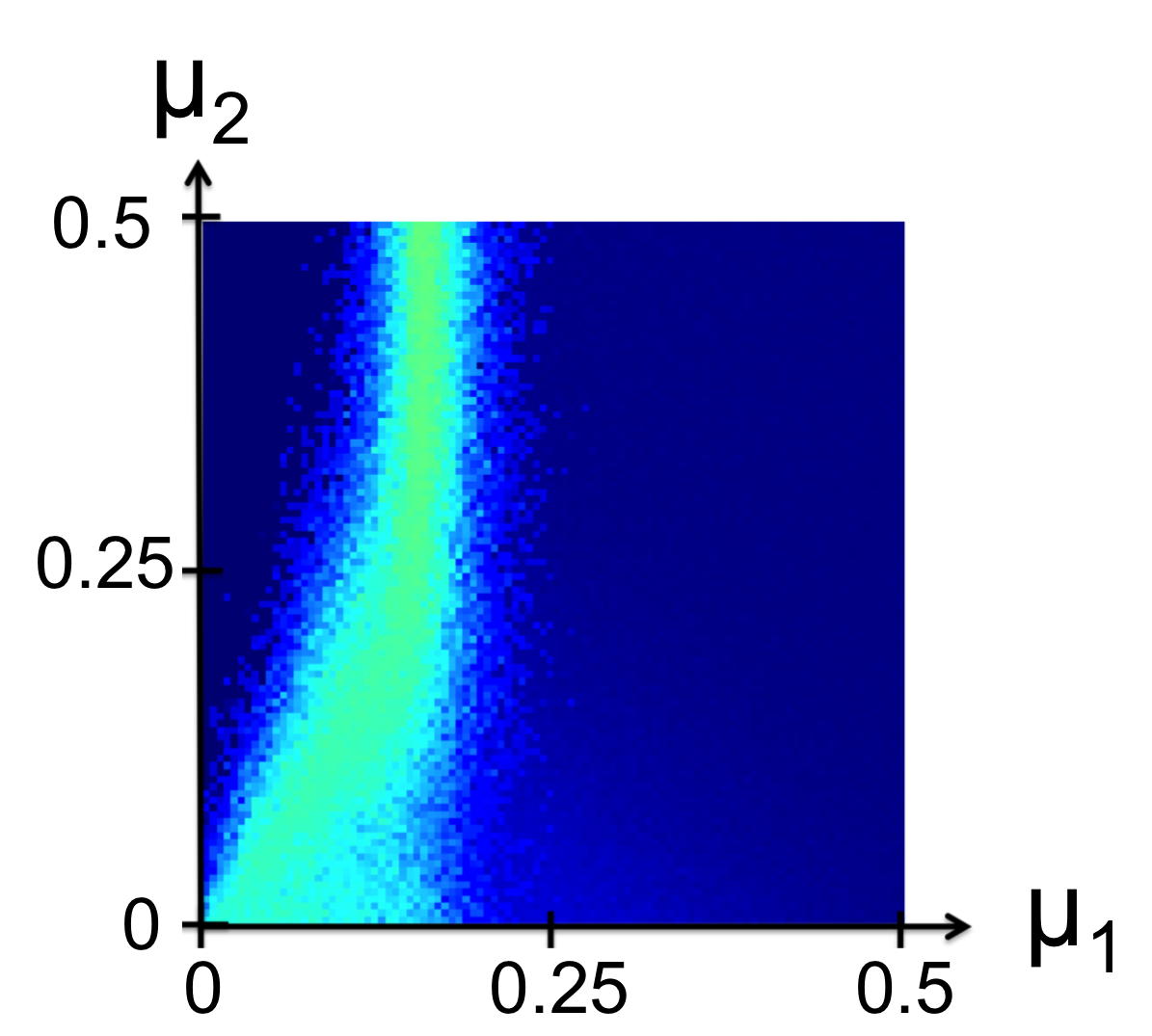} &
\includegraphics[height=1.3in]{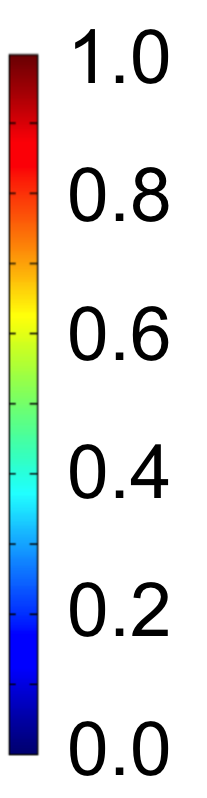}
\\
\\
\\
\multicolumn{3}{c}{Symmetric Normalized Laplacian $\boldsymbol{L_s}$: Normalized Cut} \\
\includegraphics[width=1.6in]{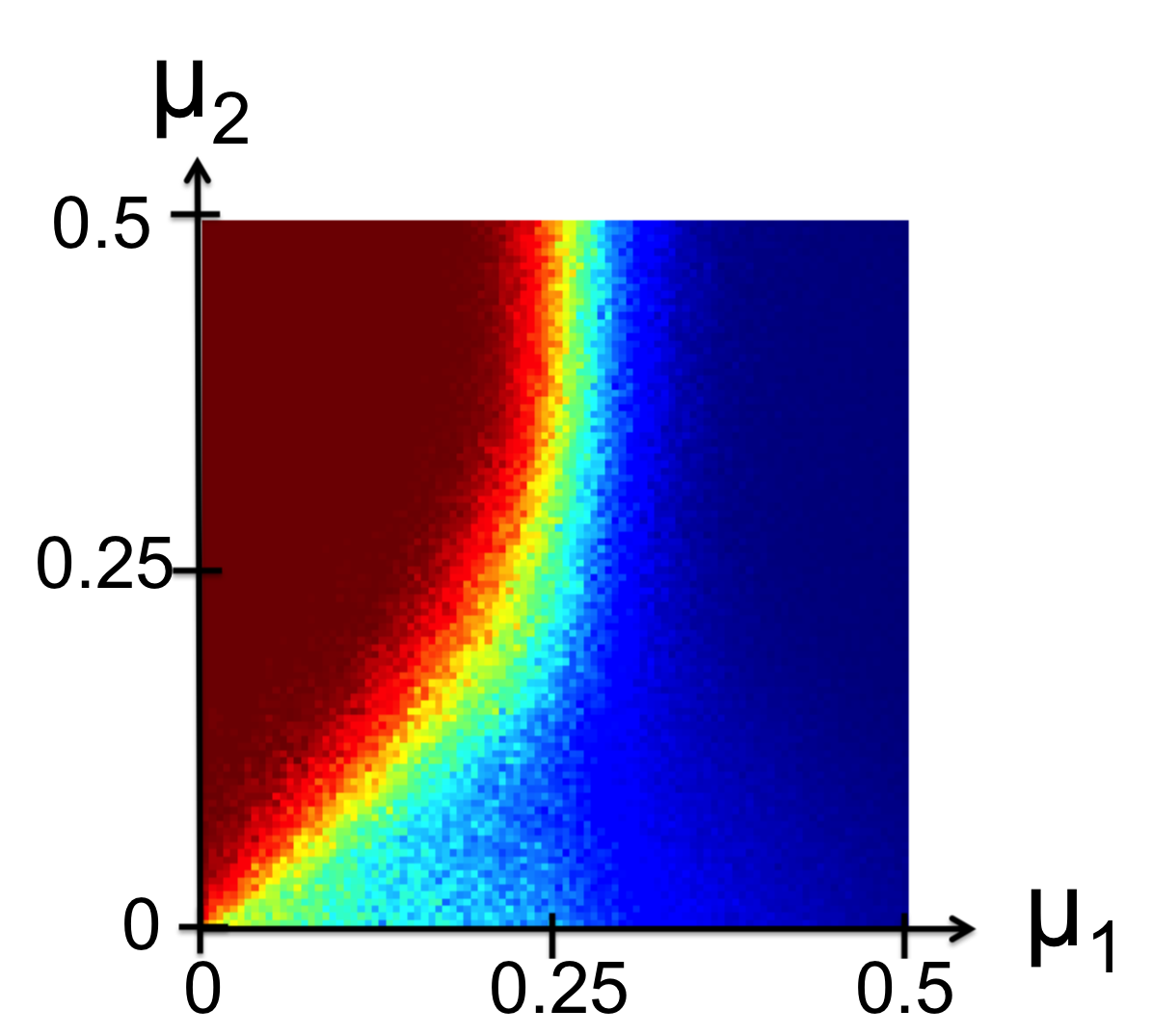} &
\includegraphics[width=1.6in]{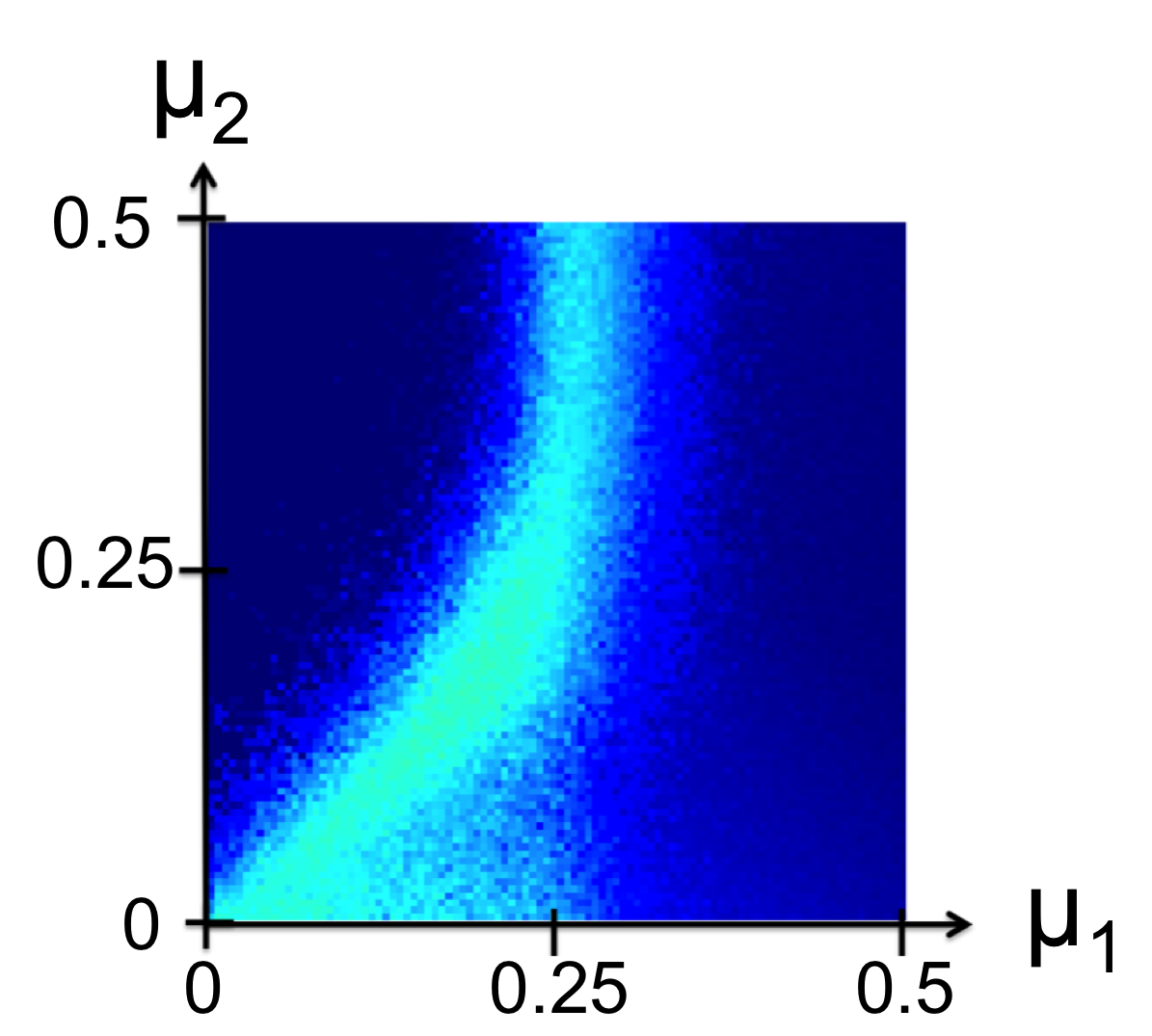} &
\includegraphics[height=1.3in]{color_scale3_3c.png}
\\
\\
\\
\multicolumn{3}{c}{Replicator $\boldsymbol{R}$: Normalized Cut (reweighted)} \\
\includegraphics[width=1.6in]{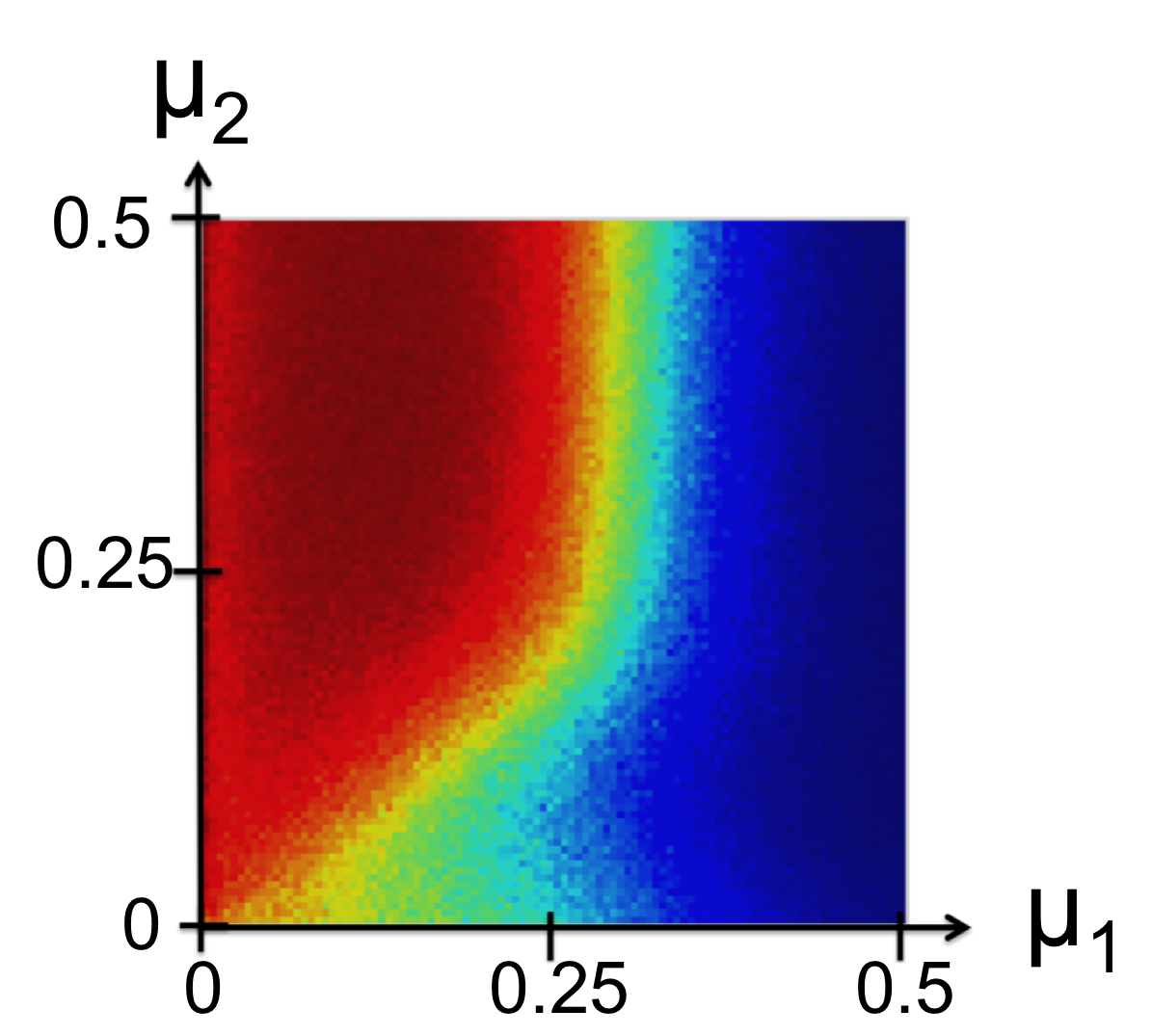} &
\includegraphics[width=1.6in]{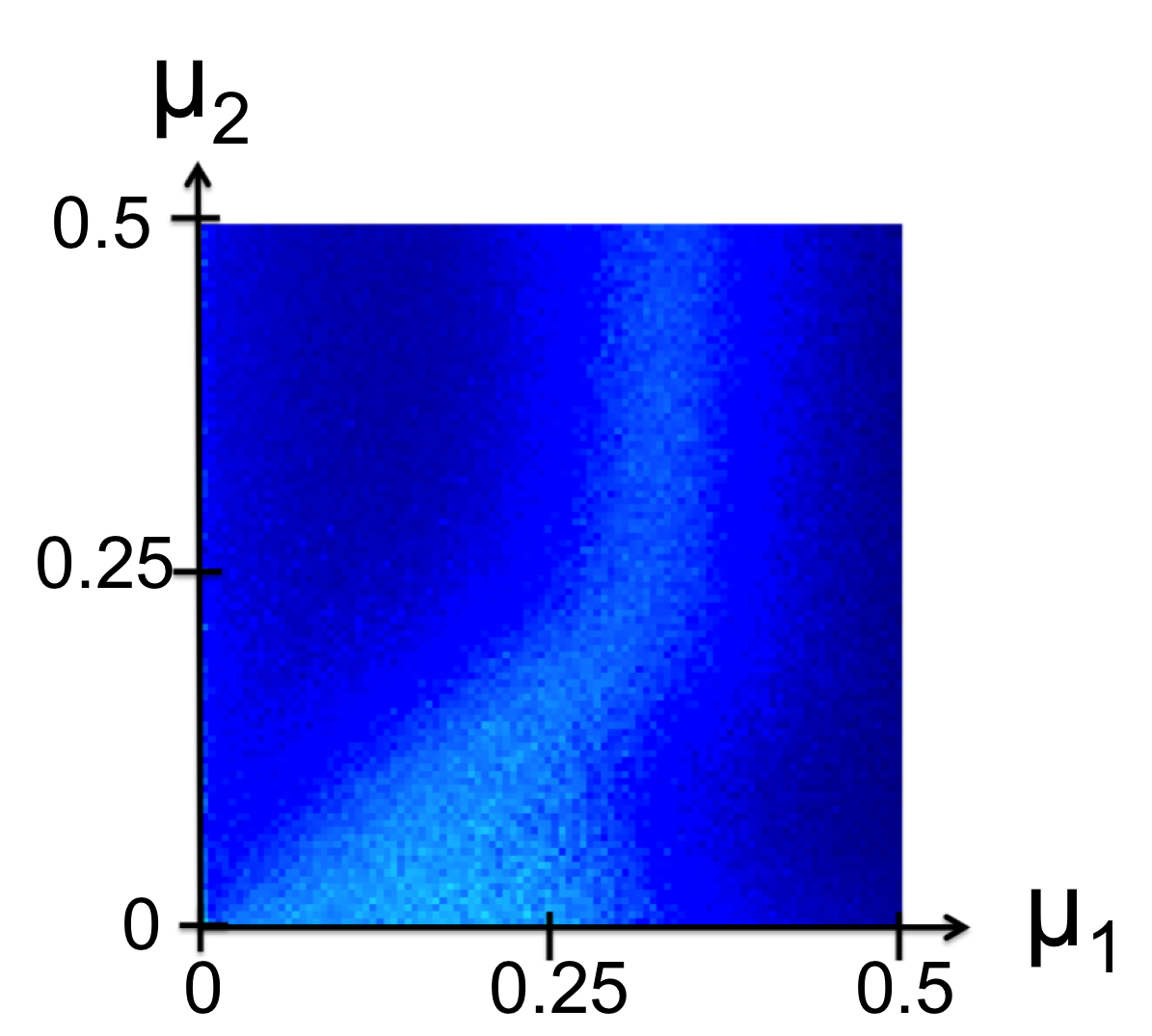} &
\includegraphics[height=1.3in]{color_scale3_3c.png}
\end{tabular}
\caption{ NMI scores for minimizing the respective operators' quality measure.  Each pixel represents the average (left) or standard deviation (right) NMI score across 100 runs for fixed  $(\mu_1, \mu_2)$. \label{fig:individualMetric}}
\end{figure}

We partition each benchmark graph using $\boldsymbol{L}$, $\boldsymbol{L}_s$, and $\boldsymbol{R}$ by minimizing their respective quality measures. To evaluate the resulting partitions, we use the Normalized Mutual Information (NMI) measure~\cite{Danon2005}, which compares the partition to the ground truth communities. When the value of this measure is 1.0, the partitioning method has successfully recovered the underlying community structure.
We calculate the average and standard deviation of the NMI scores for a fixed set of parameters and display the results in Figure~\ref{fig:individualMetric}.

As the proportion of a node's edges that connect to individuals in the opposite community, $\mu_1$, increases, it becomes more difficult to divide the network into the correct communities. We find that $\boldsymbol{L}$ and $\boldsymbol{L}_s$ give better results when $\mu_1$ is small (very few links between the two communities).  As $\mu_1$ increases, $\boldsymbol{R}$ dominates with a higher NMI score.  Additionally, $\boldsymbol{R}$  has the lowest standard deviation of the three operators, indicating a consistent performance in identifying the underlying communities.

\section{Conclusion}

Spectral partitioning traditionally uses the graph Laplacian. In this paper, we have introduced a method for spectral partitioning using the replicator, an operator describing epidemic diffusion on graphs.  We have shown that this operator is equivalent to the symmetric normalized Laplacian on a different graph, where edges are reweighted according to the eigenvector centrality measure.  By reweighting the edges, a higher weight is placed on globally important nodes.  Thus, this method tends to preserve cliques and other dense clusters.

We have introduced a  spectral bisection approach based on the componentwise ratio of the second to the first eigenvector of $\boldsymbol{R}$, choosing the partition by splitting the sorted vector so as to minimize an appropriate quality measure.  Comparing the performance of different methods on synthetic graphs with known community structure, we have shown that spectral partitioning using the  replicator is better able to recover the underlying community structure, especially in cases where more edges between the two macro communities make it more difficult for the Laplacian and symmetric normalized Laplacian to identify  communities.  By reweighting the edges using eigenvector centrality, the replicator assigns more importance to central nodes.  Thus, the edges that pass between clusters are given less influence if they do not link nodes of high centrality.  By limiting the cuts to influential edges, the method leads to a more accurate reconstruction of the community structure.

\subsection*{Acknowledgements}
The authors are grateful to Arjuna Flenner, Yves van Gennip, and Blake Hunter for many instructive conversations and suggestions. KL and RG are also greatly indebted to Shanghua Teng, whose insights and enthusiasm continue to inspire them.  This work has been funded by the Air Force Office of Scientific Research under contracts FA9550-10-1-0569  and FA9550-10-1-0102, DARPA under Contract No. W911NF-12-1-0034, the National Science Foundation under grant 1217605, and the Department of Energy Office of Science Advanced Computing Research (ASCR) program in Applied Mathematics.

%\bibliographystyle{revtex}%abbrv}
%\bibliography{references}

\end{document}